\documentclass[acmsmall]{acmart} 
\usepackage[utf8]{inputenc} 
\usepackage[T1]{fontenc}    
\usepackage{hyperref}       
\usepackage{url}            
\usepackage{booktabs}       
\usepackage{amsfonts}       
\usepackage{nicefrac}       
\usepackage{microtype}      
\usepackage{xcolor}         
\usepackage{graphicx}
\usepackage{amsmath} 
\usepackage{multirow} 
\usepackage{cleveref}
\usepackage{enumitem}
\usepackage{pgfplots}
\usepackage{tikz}
\usepackage{xcolor}
\usepackage{amsmath}
\usepackage{wrapfig}
\usepackage{algorithmic}
\usepackage{subcaption}
\usepackage{xspace}
\usepackage{tcolorbox}
\usepackage[linesnumbered,ruled,vlined]{algorithm2e}
\usepackage{listings}
\renewcommand\footnotetextcopyrightpermission[1]{} 
\newcommand{\sysname}{{\sc ParCleanse}\xspace}

\AtBeginDocument{%
  }


\newcommand{\todoc}[2]{{\textcolor{#1}{\textbf{#2}}}}

\definecolor{applegreen}{rgb}{0.55, 0.71, 0.0} 

\definecolor{light-gray}{gray}{0.7}

\makeatletter
\newcommand*{\textoverline}[1]{$\overline{\hbox{#1}}\m@th$}
\makeatother

\newcommand\blackcircle[1]{%
  \tikz[baseline=(X.base)] 
    \node (X) [draw, shape=circle, inner sep=0, scale=0.8, fill=darkgray, text=white] {\strut #1};%
}

\newcommand\whitecircle[1]{%
  \tikz[baseline=(X.base)] 
    \node (X) [draw, shape=circle, inner sep=0, scale=0.8, fill=white, text=black] {\strut #1};%
}

{\theoremstyle{definition}}
{\theoremstyle{definition}}
{\theoremstyle{definition}}
{\theoremstyle{definition}\newtheorem{definition}{Definition}}
{\theoremstyle{definition}\newtheorem{example}{Example}}

\renewcommand{\todoc}[2]{\relax}

\newboolean{showchanges}
\setboolean{showchanges}{false} 

\newcommand{\change}[1]{%
    \ifthenelse{\boolean{showchanges}}%
        {\textcolor{blue}{#1}}
        {#1}
}

\newcounter{finding}

\newcommand{\distance}{8pt}
\pgfplotsset{compat=1.18}

\begin{document}

\title{Validating Network Protocol Parsers with Traceable RFC Document Interpretation}

\author{Mingwei Zheng}
\orcid{0009-0003-6032-6045}
\affiliation{%
  \institution{Purdue University}
  \city{West Lafayette}
  \country{USA}
}
\email{zheng618@purdue.edu}

\author{Danning Xie}
\orcid{0000-0002-4359-4625}
\affiliation{%
  \institution{Purdue University}
  \city{West Lafayette}
  \country{USA}
}

\email{xie342@purdue.edu}

\author{Qingkai Shi}
\orcid{0000-0002-8297-8998}
\affiliation{%
  \institution{Nanjing University}
  \city{Nanjing}
  \country{China}
}
\email{qingkaishi@nju.edu.cn}

\author{Chengpeng Wang}
\orcid{0000-0003-0617-5322}
\affiliation{%
  \institution{Purdue University}
  \city{West Lafayette}
  \country{USA}
}
\email{wang6590@purdue.edu}

\author{Xiangyu Zhang}
\orcid{0000-0002-9544-2500}
\affiliation{%
  \institution{Purdue University}
  \city{West Lafayette}
  \country{USA}
}
\email{xyzhang@cs.purdue.edu}

\begin{abstract}
  Validating the correctness of network protocol implementations is highly challenging due to the oracle and traceability problems. The former determines when a protocol implementation can be considered buggy, especially when the bugs do not cause any observable symptoms. The latter allows developers to understand how an implementation violates the protocol specification, thereby facilitating bug fixes.
  Unlike existing works that rarely take both problems into account,
  this work considers both and provides an effective solution using recent advances in large language models (LLMs).
  Our key observation is that network protocols are often released with structured specification documents, a.k.a. RFC documents, which can be systematically translated to formal protocol message specifications via LLMs.
  Such specifications, which may contain errors due to the hallucination of LLMs, are used as a quasi-oracle to validate protocol parsers, while the validation results in return gradually refine the oracle.
  Since the oracle is derived from the document, any bugs we find in a protocol implementation can be traced back to the document, thus addressing the traceability problem.
  We have extensively evaluated our approach using nine network protocols and their implementations written in C, Python, and Go.
  The results show that our approach outperforms the state-of-the-art
  and has detected 69 bugs, with 36 confirmed.
  The project also demonstrates the potential for fully automating software validation based on natural language specifications, a process previously considered predominantly manual due to the need to understand specification documents and derive expected outputs for test inputs.
\end{abstract}


\maketitle

\section{Introduction}
Network protocols play a key role in the Internet of Things era as they define how devices worldwide connect to and communicate with each other.
As essential components in network protocol implementations, network protocol parsers parse and validate network messages,
which ensures network messages follow specific syntactic and semantic rules, thus preventing invalid or malicious data from disrupting system operations or compromising security.
Despite their importance, building high-quality network protocol parsers is challenging and error-prone~\cite{everparse, Hardening}.
According to MITRE, input (e.g., network messages) validation issues~\cite{CWE-20} are among the top four in the CWE Top 25 Most Dangerous Software Weaknesses~\cite{cwe25}, emphasizing the critical risks that improperly handled inputs pose to system security and reliability.

\begin{figure}[t]
  \centering
 \includegraphics[clip=true,trim=0mm 0.5mm 0mm 0mm,width=\linewidth]{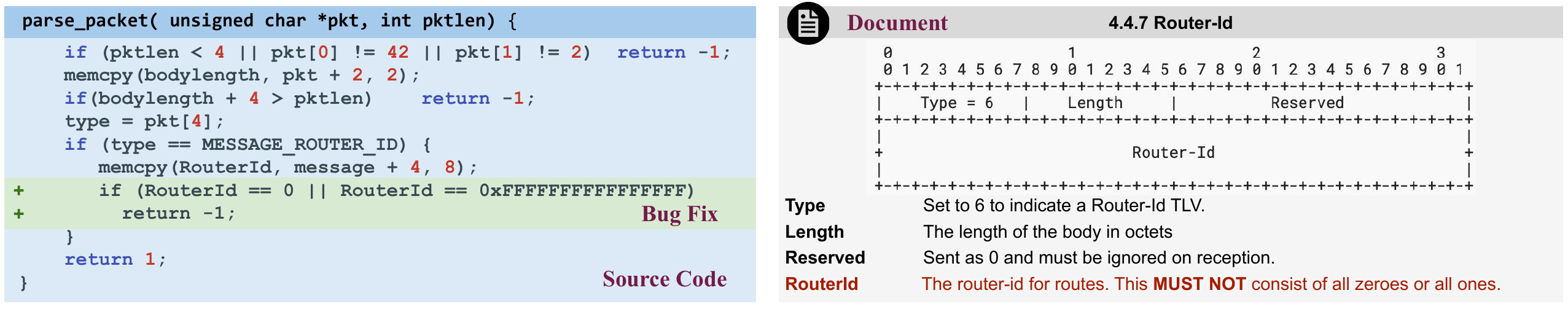}
  \caption{A bug detected by \sysname, its fix, and the corresponding documentation.
  }
  \label{fig:motivation_code}
\end{figure}

\vspace{1mm}
\noindent
\textbf{Existing Works.} 
Our work targets input validation bugs in network protocol parsers, where the parser incorrectly accepts invalid packets or rejects valid packets. 
Unfortunately, existing test oracles are insufficient to thoroughly detect such bugs because many of them are silent, not causing obvious runtime symptoms (e.g., crashes) or violating other well-established properties (e.g., memory-safety and data-privacy).
As a result, conventional fuzzing~\cite{Msrsage,boofuzz} and static analyzers~\cite{shi2018pinpoint, klee} that rely on non-protocol-specific oracles can hardly detect them.
Model checking~\cite{MusuvathiE04, DiazCRP04} constructs protocol-specific oracles to detect these bugs. 
However, formal specifications are often missing. Manually constructing them is time-consuming since network protocols are often described in natural language (e.g., in RFC documents). 
To address these oracle issues, differential analysis techniques, including both static differential analysis~\cite{Pardiff, SymCerts} and dynamic differential analysis~\cite{reen2020dpifuzz, xDiff}, identify bugs by comparing multiple implementations of the same protocol.
While they are effective in many cases, they fail to detect bugs shared by multiple implementations.
For instance, the bug in \Cref{fig:motivation_code} actually exists in multiple Babel implementations, including FRRouting Protocol Suite~\cite{frr} and Jech/Babel~\cite{jech},
where differential analysis techniques become ineffective.

\vspace{1mm}
\noindent
\textbf{Our Approach.} 
A popular software validation method is to extract testable properties from specification documents and then construct inputs to test these properties. The key challenge is deriving the expected outputs for given inputs based solely on the documents, which often requires substantial manual effort.
Our paper presents \sysname, which uses Large Language Models (LLMs) to automate this process. In particular, it extracts formal protocol specifications from RFC documents and derives inputs together with the expected outputs from the specifications to validate
 parser implementations.
\sysname generates a set of valid packets (conforming to the format) and invalid packets (violating the format) based on the LLM-extracted protocol formats to test whether target parsers accept valid packets and reject invalid ones.
If a parser deviates from this expected behavior, it may indicate a potential inconsistency between the specification and its implementation.
Since RFC documents are widely accepted as network protocol standards, \sysname overcomes the limitations of differential analysis, which cannot detect bugs shared across multiple implementations.
Furthermore, the LLM-based format extraction process is highly automated, reducing the human effort to generate protocol-specific oracles.

However, many RFC documents are lengthy and complex, causing substantial LLM hallucinations. As such, extracted protocol formats may contain errors, leading to misinterpretation of parser behavior. 
For example, an incorrect field constraint in the extracted format could cause us to mistakenly flag a parser bug if the parser accepts a packet that violates the constraint.
To mitigate LLM hallucinations, our approach incorporates two key designs:
\vspace{-2mm}
\begin{itemize}[leftmargin=0.3cm]
    \item First, our approach uses a divide-and-conquer strategy to systematically decompose an RFC document into smaller and manageable sections while preserving their structural relationships within a knowledge graph called the DocTree. 
This decomposition mitigates LLM hallucinations across large contexts and enables precise extraction of subformats, which are then combined into a complete protocol format with the hierarchical guidance of the DocTree.

    \item Second, our approach features traceable inconsistency identification: when an inconsistency between the LLM-extracted format and parser behavior is detected, it is traced back to the relevant sections of the RFC document. This traceability supports an additional validation step to determine whether the inconsistency stems from the LLM's hallucination or a parser implementation bug.
\end{itemize}

To thoroughly validate parser implementations, \sysname performs both field-level and structure-level mutations (referred to as \textbf{property-level mutations}) to generate comprehensive test cases. 
It generates both positive and negative inputs for each property, allowing thorough validation of parser implementations by verifying each protocol property against its specification.

\smallskip
\noindent
\textbf{Contribution.} 
In summary, we make the following contributions.

\begin{itemize}[leftmargin=0.3cm]
   \item We propose a novel validation approach to detect bugs in network protocol parsers by ``comparing\text{''} them with protocol formats extracted from RFC documents. 
    \begin{itemize}[leftmargin=0.3cm]
        \item It conducts a divide-and-conquer format extraction to interpret the official RFC documents of network protocols to precise and complete network protocol format specifications.

        \item It features fine-grained property-level input mutations to thoroughly test parser implementations guided by the extracted specifications.

        \item It leverages a traceable inconsistency identification technique, allowing any identified inconsistencies to be traced back to the original specification for a more accurate diagnosis.
    \end{itemize}\smallskip
    
    \item We implement our approach as a prototype tool, \sysname,\footnote{\sysname is publicly available at \url{https://github.com/zmw12306/ParCleanse}.} and evaluate it on nine network protocols implemented in C, Python, and Go. The experimental results show that \sysname effectively extracts protocol formats from RFCs, achieving 100\% precision and recall for message types and 99\% precision and 95\% recall for field names, outperforming the state-of-the-art LLM-based method, ChatAFL.
   \sysname identifies 69 bugs, with 36 confirmed, outperforming the state-of-the-art protocol parser testing tool, ParDiff.
\end{itemize}

\section{Motivation}\label{sec:motiv}
In this section, we first present a real-world bug detected by \sysname (\Cref{subsec:bug}) and illustrate the limitations of existing methods (\Cref{subsec:limitations}). 
We then discuss the inherent challenges of the problem and introduce the design of our technique (\Cref{subsec:approach}).

\subsection{A Real-World Example}\label{subsec:bug}
\Cref{fig:motivation_code} shows a buggy code snippet of the Babel network routing protocol (from the FRRouting Protocol Suit~\cite{frr})  and its RFC document.
The document specifies the Babel protocol format, which includes four fields: \texttt{Type}, \texttt{Length}, \texttt{Reserved}, and \texttt{RouterId}.
In RFC documents, the width of each field in the structure table indicates its byte length. In this example, \texttt{Type} is one byte, while \texttt{RouterId} is eight bytes.
This document also restricts \texttt{RouterId} from being all zeroes (\texttt{0}) or all ones (\texttt{0xFFFFFFFFFFFFFFFF} for eight bytes), as these could lead to ambiguities in Babel's routing strategies. The check on \texttt{RouterId} helps Babel maintain stability and prevent routing loops, especially in dynamic or frequently changing network topologies.
However, this validation is missing in the buggy implementation.
Malicious attackers could exploit this oversight to introduce packets with an invalid \texttt{RouterId} into the network and disrupt the network, causing instability in path calculation and failure to properly update routing tables.

\subsection{Limitations of Existing Work}\label{subsec:limitations}

Despite the severity of this issue, existing techniques struggle to detect the bug.
A key limitation of existing work is the lack of high-quality oracles.
Traditional static analyzers (e.g., KLEE~\cite{klee}, Pinpoint~\cite{shi2018pinpoint}) and conventional fuzzing techniques (e.g., SAGE~\cite{Msrsage}, BooFuzz~\cite{boofuzz}) rely on general oracles, such as the violations of safety properties and the abnormal behaviors in the runtime, to detect and trigger bugs, respectively.
Existing LLM-based testing approaches like ChatAFL~\cite{chatafl} and Fuzz4All~\cite{Fuzz4All} also rely on crash-based oracles.
However, the bug in \Cref{fig:motivation_code} does not violate such general properties,
but rather protocol-specific properties causing silent system state corruptions.

Differential analysis tools like ParDiff~\cite{Pardiff} and DPIFuzz~\cite{reen2020dpifuzz} partially address the need for protocol-specific oracles by comparing multiple implementations of the same protocol. However, these approaches require at least one correct implementation to flag inconsistencies. 
For the bug in \Cref{fig:motivation_code}, alternative implementations (e.g., Jech/Babel~\cite{jech}) also miss checking the specific condition. As such, differential analysis cannot detect the bug. 
Additionally, although model checking could theoretically verify protocol-specific properties, it requires a formal protocol specification, which is missing for Babel. Unfortunately, constructing a formal model from Babel’s natural language RFC document would require substantial manual effort.

\begin{figure}[t]
  \centering
 \includegraphics[clip=true,trim=0mm 0.5mm 0mm 0mm,width=\linewidth]{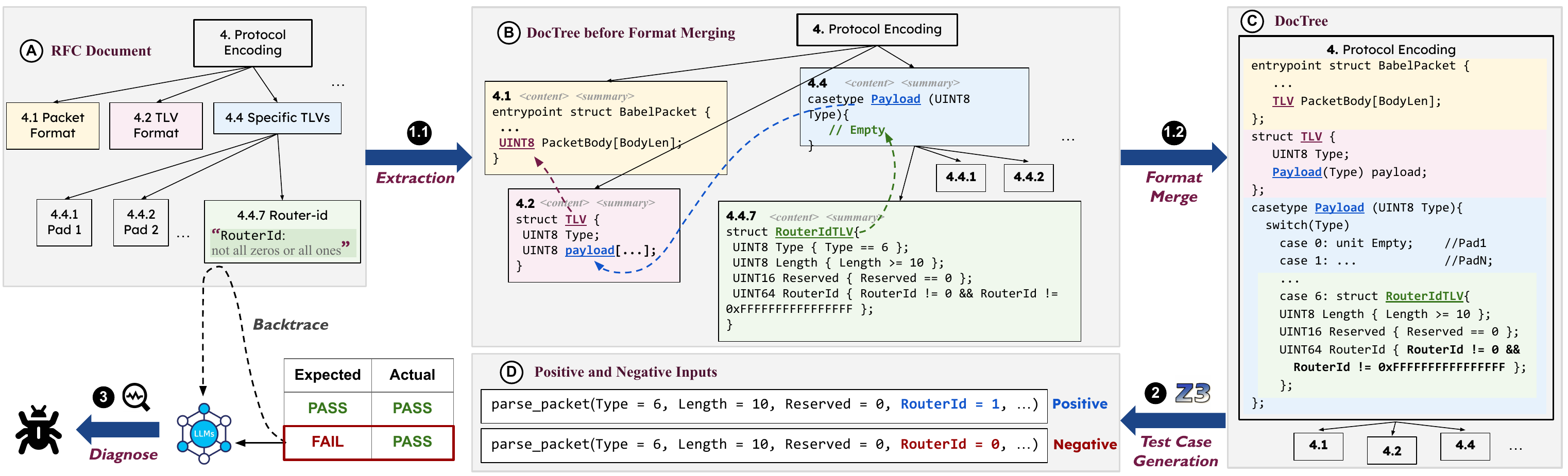}
  \caption{ \sysname extracts specifications from documents to build DocTree. Dashed arrows in \protect\whitecircle{B} indicate hierarchical relationships for forming the protocol format in \protect\whitecircle{C}. \sysname then generates test cases to detect inconsistencies and backtraces the relevant document section for LLM diagnosis.}
  \label{fig:motivation_process}
\end{figure}
\subsection{Our Approach in a Nutshell}\label{subsec:approach}
We propose \sysname, which leverages LLMs to automatically generate oracles (i.e., network protocol format specifications) from protocol documents and create fine-grained test cases to effectively discover bugs with a low false positive rate.
The extracted oracles have two main benefits: (1) covering all formats described in the document, allowing \textit{comprehensive} testing of protocol parsers; and (2) providing \textit{traceability}, which enhances bug understanding, reduces false positives, and facilitates fixes.
We will walk through the motivating example, introduce the three phases of \sysname, and outline the challenges and solutions at each step.

\vspace{1.5mm}
\noindent\textbf{Phase 1: Protocol Format Extraction  (\Cref{section: stage1}).}
An intuitive approach is to extract formal specifications directly from RFC documents, as they define expected input and output behaviors with high quality. 
However, RFC documents are often long and complex, ranging from tens to hundreds of pages. 
Feeding the entire document into an LLM to automatically generate protocol format specifications (or formats for simplicity) often leads to incomplete or incorrect formats due to LLMs’ inherent hallucinations.
A more effective approach is to divide the document into smaller pieces and prompt the LLM to generate the format for each piece. However, a challenge arises:

\vspace{1mm}
\noindent
\fcolorbox{black}{gray!10}{
\parbox{0.98\columnwidth}{
    \textbf{Challenge 1:}
    While dividing a lengthy RFC document into smaller pieces can reduce errors from LLM outputs, these pieces are often interdependent, making it difficult to directly combine inferred sub-protocol formats.
}}
\vspace{1mm}

We implement a divide-and-conquer strategy using a data structure, DocTree, to capture hierarchical relationships within RFC documents. 
As shown in \Cref{fig:motivation_process}, we begin by dividing the RFC 8966 for the Babel protocol into sections (\whitecircle{A}). In step \blackcircle{1.1}, the LLM summarizes each section and generates the corresponding protocol format sub-specification.
Based on the segmentation, we construct an initial DocTree (\whitecircle{B}) that mirrors the document’s table of contents, treating each section as a node. However, the hierarchical structure in the table of contents doesn’t align with the protocol’s message structure. For example, while \textit{``section 4.1\text{''}} (packet format) and \textit{``section 4.2\text{''}} (TLV format) appear at the same level in the table of contents, they actually have a nested relationship. Specifically, \textit{``4.1\text{''}} should be the parent of \textit{``4.2\text{''}}, as \textit{``4.2\text{''}} represents a subset of the network packet outlined by \textit{``4.1\text{''}}.
To correct this misalignment, we prompt the LLM in step \blackcircle{1.2} to refine the DocTree hierarchy based on the relationships between sections as shown with the dashed arrows, producing a structure that accurately reflects the protocol’s format. This refined DocTree enables us to generate a coherent, hierarchical protocol format as shown in \whitecircle{C}.

\vspace{1.5mm}
\noindent\textbf{Phase 2: Test Case Generation (\Cref{section: stage2}).}
Given the protocol formats with field constraints in \whitecircle{C}, we generate test cases based on the constraints. While solvers like Z3 can efficiently produce both valid and invalid inputs that satisfy or violate these constraints, using the full set of constraints alone does not sufficiently explore the input space, nor does it guarantee diversity in bug discovery.

This limitation arises because the generated tests do not isolate specific fields or individual constraints.
As a result, although direct test generation based on the whole specification may trigger some exceptions, it tends to uncover only a small set of bugs, leaving large portions of the input space untested and essential format properties (field-level properties and structure-level properties) unchecked (see the ablation studies in \Cref{sec:ablation_bug} for further discussions).

\vspace{1mm}
\noindent
\fcolorbox{black}{gray!10}{
\parbox{0.98\columnwidth}{
    \textbf{Challenge 2:}
    Naively generating inputs based on the whole protocol specification is insufficient to validate each format property and exhaustively explore parser behaviors.
}}
\vspace{1mm}

\sysname addresses this challenge with a fine-grained test-generation strategy. In step \blackcircle{2}, we produce both positive and negative inputs that conform and violate the individual constraints (\whitecircle{D}), respectively. A negative input violates only a single format property. If the target parser accepts both types, it indicates a lack of enforcement of the property, highlighting inconsistencies between the specification and the implementation. This systematic, property-level mutation provides a comprehensive protocol validation.

As shown in \whitecircle{D}, to validate the format property for the field \texttt{RouterId}, we generate two network messages: one with \texttt{RouterId = 1} conforming the constraint, and the other with \texttt{RouterId = 0} violating it. In the negative test case, all other fields are identical to the positive case. However, both test cases are accepted by the parser, indicating an inconsistency where the parser fails to enforce the \texttt{RouterId} constraint specified by the protocol format.

\vspace{1.5mm}
\noindent\textbf{Phase 3: Inconsistency Identification (\Cref{section: stage3}).}
Even when an inconsistency is detected, inaccuracies in the extracted specification may cause false positives, posing this challenge:

\vspace{1mm}
\noindent
\fcolorbox{black}{gray!10}{
\parbox{0.98\columnwidth}{
    \textbf{Challenge 3:}
    How can we distinguish real inconsistencies from false positives caused by incorrect specifications?
}}
\vspace{1mm}

Thus, when an inconsistency arises between the extracted specification and the parser implementation, it is crucial to determine whether it stems from an implementation bug or an error in the LLM-extracted format. \sysname enables this by providing complete traceability from the triggered inconsistency back to the exact RFC section where the relevant property is defined. 
By linking each format property with its source RFC section during format extraction (Phase 1), we establish traceability, allowing an accurate diagnosis of potential format violations in parsers.

In the motivating example, we generate the negative input by intentionally violating the constraint for \texttt{RouterId} as specified in \textit{``section 4.4.7\text{''}}. To diagnose this inconsistency, we retrieve \textit{``section 4.4.7\text{''}} through ``backtrace\text{''} (indicated by the dashed arrow) from the RFC document and prompt the LLM to analyze the discrepancy.
To enhance accuracy, we use chain-of-thought prompting, which requires the LLM to explain its reasoning steps, thereby reducing errors and improving decision reliability. In this case, the LLM concludes that it is due to a bug in the parser implementation. We hence report the bug (i.e., missing checks for RouterId not being all zeros or all ones). On the other hand, if the LLM concludes that the inconsistency is due to the incorrect extracted format, it will refine the format accordingly to improve the testing effectiveness.
\section{Problem Formulation}\label{sec:problem}

This section introduces key preliminaries, including the formal definitions of network packets and their format syntax, followed by a formal statement of the problem addressed in this paper.

\subsection{Protocol Packet and Its Format Syntax}\label{syntax}

Network protocols establish the standards of data transmission and interpretation for the communication between devices.
Typically, data is encoded and transmitted as a sequence of bytes, referred to as a packet.
Similar to object fields in memory, the bytes located in a consecutive segment of a protocol packet can indicate a specific unit of information,
such as message types and message contents.
Concretely, a packet should consist of the following five key elements:

\vspace{-1mm}
\begin{itemize}[leftmargin=0.6cm]
\item \textbf{Message Types:} The different formats a protocol use for various kinds of messages.
\item \textbf{Field Names:} The names of individual data fields within a specific type of message.

\item \textbf{Field Types:} The data types of fields (e.g., byte, bit, struct, and array) and their sizes, which can be either fixed or variable (i.e., depending on some other field).

\item \textbf{Independent Constraints:} Restrictions on individual fields, such as numeric ranges or fixed values, that do not rely on other fields.

\item \textbf{Dependent Constraints:} Constraints that involve relationships across multiple fields.
\end{itemize}

\begin{wrapfigure}[14]{r}{0.58\textwidth}
\vspace{-5mm}
\centering
\scalebox{1.0}{
\begin{minipage}{0.58\textwidth}
\centering
\begin{figure}[H]
\scriptsize
\centering
\[
\begin{aligned}
	\textbf{Packet} \quad & \mathit{p} \in \textsf{Struct Type}\\
	\textbf{Type} \quad & \textsf{Type} := \textsf{pType} \ | \ \textsf{ArrayType} \ | \ \textsf{CaseType} \ | \ \textsf{StructType} \\
	\textbf{Identifier} \quad & \mathit{identifier} \in \textsf{String} \\
	\textbf{Primitive Type} \quad & \textsf{pType} := \textsf{UINT8} \ | \ \textsf{UINT16} \ | \ \textsf{UINT32} \ | \ \textsf{UINT64} \ | \ \cdots \\
	\textbf{Array Type} \quad & \textsf{ArrayType} := \textsf{Type}[\mathit{const}] \ | \  \textsf{Type}[\mathit{f}(\mathit{identifier}^{+})] \\
	\textbf{Case Type} \quad & \textsf{CaseType} := \textsf{switch}(\mathit{identifier}) \ \{\textsf{case}  \ \mathit{const}: \textsf{Type} \}^{+} \\
	\textbf{Struct Type} \quad & \textsf{StructType} := \textsf{struct} \ \mathit{identifier} \  \{ \mathit{field}^{+} \} \  (\mathit{f}(\mathit{field}^{+}) \odot 0)^{*}\\
	\textbf{Field} \quad & \mathit{field} := \textsf{Type} \ \mathit{identifier} \\
	\textbf{Constant} \quad & \mathit{const} \in \textsf{UINT64}\\
	\textbf{Function} \quad & \mathit{f} \in \textsf{ArithFunction}\\
    \textbf{Cmp Operator} \quad & \odot := \ \geq \ | \ \leq \ | \ > \ | \ < \ | = \ | \ \neq\\
\end{aligned}
\]
\caption{The format syntax of protocol packets}
\label{fig:syntax}
\end{figure}
\end{minipage}
}  
\end{wrapfigure}

To ensure correct interpretation, these elements must be organized in a specific form defined in Figure~\ref{fig:syntax}.
Each packet is a \textsf{Struct Type} object, aggregating multiple fields, each defined by a \textsf{type} and \textsf{identifier}, with optional constraints.
Apart from the \textsf{Struct Type}, a \textsf{type} in the protocol format may be a \textsf{Primitive Type}, \textsf{Array Type}, or \textsf{Case Type}. \textsf{Primitive Type} (e.g., \textsf{UINT8}, \textsf{UINT16}) defines fixed-size fields.
\textsf{Array Type} represents a sequence of elements of the same type.
The length of an Array-Typed object can be constant or determined by an arithmetic expression relating to some other Primitive-Typed fields.
\textsf{Case Type} defines different possible layouts for a field or group of fields, depending on the value of a control field.
This structure enables different formats for various message types, assigning a unique structure to each type based on its specific functionality.
Without loss of generality, we suppose that the constraints upon the fields in a Struct-Typed object are arithmetic constraints,
which are in the form of $\mathit{f}(\mathit{field}^{+}) \odot 0$.
Here $\mathit{f}$ is an arithmetic function and $\odot$ is a comparison operator.

\noindent\textbf{Example.}
In \Cref{fig:motivation_process}, \whitecircle{C}:
"$\texttt{UINT8}$" is a \textsf{Primitive Type}.
$\texttt{TLV}$ is a \textsf{Struct Type} with two fields: \texttt{Type} and \texttt{Payload}, arranged sequentially in memory.
`$\texttt{TLV} \ \texttt{PacketBody[BodyLength]}$' defines \texttt{PacketBody} as a sequence of TLVs, occupying a total of \texttt{BodyLength} bytes. 
\texttt{Payload} is a \textsf{Case Type} controlled by \texttt{Type}. For $\texttt{Type} = 0$, \texttt{Payload} is an empty struct, representing \texttt{Pad1}; for $\texttt{Type} = 6$, \texttt{Payload} represents \texttt{RouterIdTLV}.

\subsection{Problem Statement}
\begin{definition}[Network Protocol Parser]
A network protocol parser is a function \( f \) that maps a network packet \( p \) to an element in the set \( \{ \textsf{pass}, \textsf{fail}, \textsf{crash} \} \). Specifically, \( f(p) = \textsf{pass} \) or \( \textsf{fail} \) indicates whether the packet conforms to the protocol format, while \( f(p) = \textsf{crash} \) indicates a failure of the parser when processing \( p \).
\end{definition}

\begin{definition}[Network Protocol Document and Approximate Format]
An RFC document specifies the valid format of protocol packets in natural language. Ideally, we would derive a precise protocol format \( \mathcal{F} \) from the RFC document, a function where \( \mathcal{F}(p) = \textsf{pass} \) if a packet \( p \) is valid, and \( \textsf{fail} \) otherwise.
However, due to inaccuracies in automated format extraction, we obtain an approximate format \( \mathcal{F}' \) instead of \( \mathcal{F} \). This approximation \( \mathcal{F}' \) may not fully align with \( \mathcal{F} \), and therefore \( \mathcal{F}'(p) \) may or may not match \( \mathcal{F}(p) \) for any given packet \( p \). Consequently, observing \( f(p) \neq \mathcal{F}'(p) \) does not directly imply \( f(p) \neq \mathcal{F}(p) \), as the discrepancy could be due to inaccuracies in \( \mathcal{F}' \).

\end{definition}

Our objective is to identify packets \( p \) for which \( f(p) \neq \mathcal{F}'(p) \) and to determine if these discrepancies are due to (1) errors in the parser \( f \), or (2) errors in the extracted approximation \( \mathcal{F}' \). For case (2), we iteratively refine \( \mathcal{F}' \) to more closely align it with \( \mathcal{F} \).

\section{Design}\label{sec:design}
\begin{figure}[t]
  \centering
 \includegraphics[clip=true,trim=0mm 0.5mm 0mm 0mm,width=\linewidth]{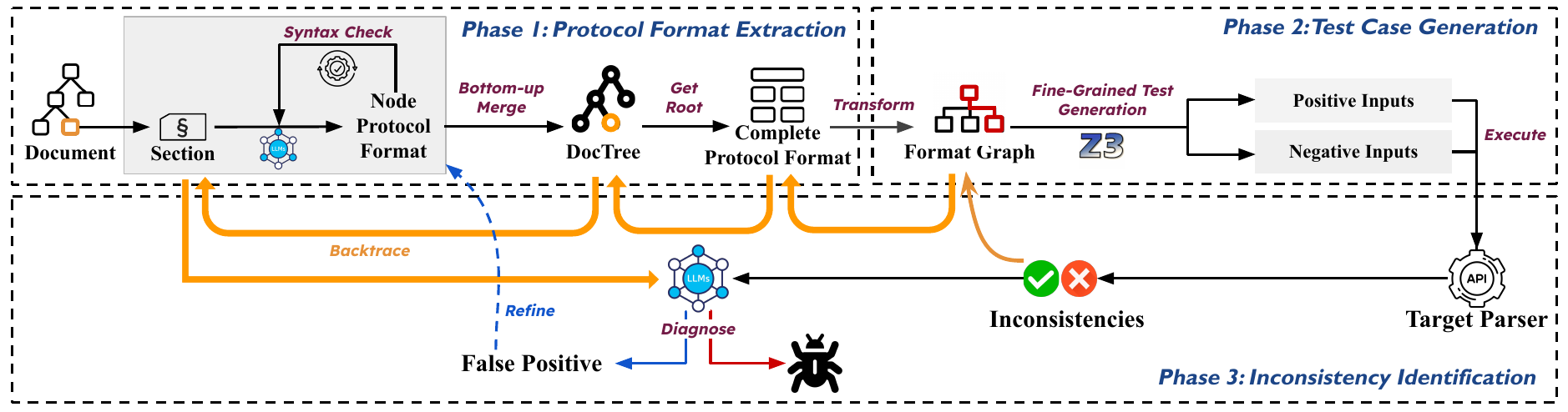}
  \caption{The pipeline of \sysname}
  \label{fig:pipeline}
\end{figure}

We present \sysname, which automatically extracts complete protocol formats from RFC documents and validates network protocol parsers with traceability-assisted inconsistency identification.
As shown in \Cref{fig:pipeline}, \sysname consists of three phases.
Specifically, Phase 1 (\Cref{section: stage1}) extracts the complete protocol format from the RFC document via divide-and-conquer. This is achieved by extracting the protocol format from each section, followed by a bottom-up merge to get the complete protocol format. The DocTree maintains a mapping between document content and extracted format to enable traceability.
Phase 2 (\Cref{section: stage2}) generates fine-grained test cases for the target parser, including both positive and negative inputs created via property-level mutation, which ensures that each protocol property is thoroughly tested against the parser. 
If an inconsistency is detected between parser executables and the extracted formats, Phase 3 (\Cref{section: stage3}) identifies the root cause by tracing back to the relevant document section and either reports a bug or diagnoses it as a false positive and refines the format.
\subsection{Phase 1: Protocol Format Extraction}\label{section: stage1}
As shown in \Cref{fig:pipeline}, Phase 1 extracts complete protocol formats from RFC documents using a divide-and-conquer approach with a hierarchical DocTree structure. Each document section is represented as a node in the DocTree, with edges capturing hierarchical relationships. Protocol formats are extracted from each node (i.e., RFC section) individually, with syntax checks conducted on generated formats. If invalid, the error is fed back to the LLM for regeneration. These formats are then merged bottom-up within the DocTree, with the root node ultimately representing the complete protocol format.
In what follows, we first introduce the DocTree structure and its initial generation in \Cref{subsec:DocTree}. 
Then, we discuss how DocTree supports the divide-and-conquer strategy for format extraction in \Cref{subsec:formatgen}. 
Finally, we discuss how traceability is maintained throughout this phase in \Cref{subsec:trace}.

\subsubsection{DocTree Initialization}\label{subsec:DocTree}

DocTree is a hierarchical representation of an RFC document, designed to reflect its structure.
It preserves relationships between different sections, essential for combining extracted formats into a complete protocol format.

\begin{definition}
A DocTree is formally defined as a tuple $\mathcal{T}_{\text{DocTree}} = (N, E)$, where: 
\begin{itemize} [leftmargin=0.6cm]
\item $N$ is a set of nodes, with each node $n_i$ defined as $n_i = (\texttt{content}, \texttt{summary}, \texttt{format})$, representing a specific RFC document section. Each node includes (1) the section content and summary and (2) the protocol format extracted from that section.
\item $E \subseteq N \times N$ is a set of directed edges. Each edge $(n_1, n_2) \in E$ represents a hierarchical relationship where $n_2$ is a subcomponent or dependent section of $n_1$. These edges preserve the document’s structure, supporting the accurate merging of section formats into a complete protocol format.

\end{itemize} 
\end{definition}

\sysname begins by constructing the initial DocTree using the RFC document's table of contents. However, sections at the same level often have implicit dependencies that are not reflected in the table of contents, which are critical for accurately merging extracted formats. 
To capture these hidden dependencies, \sysname employs a rule-based prompting method.
First, \sysname prompts LLMs to summarize each section.
Then, it re-prompts the LLMs with the summaries of sections at the same hierarchical level, asking them to identify any dependencies between them. The prompts are as follows:
\begin{tcolorbox}[colback=gray!10, colframe=gray!60, sharp corners, title=Prompt to generate section summaries, 
fonttitle=\scriptsize, 
boxsep=0.5mm, 
top=0.5mm, 
bottom=0.5mm] 
\scriptsize 
Task: Please summarize a given RFC section: \{Section\}
\end{tcolorbox}

\begin{tcolorbox}[colback=gray!10, colframe=gray!60, sharp corners, title=Prompt to identify hierarchy dependencies among sections, 
fonttitle=\scriptsize, 
boxsep=0.5mm, 
top=0.5mm, 
bottom=0.5mm] 
\scriptsize 
Task: Analyze the hierarchical structure of the following sections in an RFC document: \{Section Summaries\}\\
Instructions:
Identify Parent-Child relationships where one section provides a detailed breakdown of another.
\end{tcolorbox}
This method allows \sysname to uncover hidden dependencies, ensuring a detailed and accurate DocTree representation for later format merge.

\begin{figure*}[ht]
\centering

\begin{minipage}{0.58\textwidth} 
    \centering
    \begin{algorithm}[H]
\footnotesize
\caption{\footnotesize Protocol Formats Extraction using DocTree}
\label{alg:protocol_extraction}
\KwIn{$\mathcal{T}_{\text{DocTree}}$: DocTree, $DSL$: Protocol format syntax}
\KwOut{$F_{\text{complete}}$: Complete Protocol Format}

\SetKwFunction{FMain}{ProtocolFormatExtraction}
\SetKwFunction{FMerge}{MergeFormats}
\SetKwProg{Fn}{Function}{:}{}

\Fn{\FMain{$\mathcal{T}_{\text{DocTree}}$, $DSL$}}{
    \ForEach{$n_i \in \mathcal{T}_{\text{DocTree}}$}{ \label{line:foreach_node}
        $F_i \gets$ \texttt{LLMGenerateFormat}($n_i.\text{content}, DSL$)\; \label{line:generate_format}
        $(v, \text{errorMsg}) \gets$ \texttt{SyntaxChecker}($F_i$)\; \label{line:syntax_check}
        \While{$v == \text{False}$}{ \label{line:while}
            $F_i \gets$ \texttt{LLMSyntaxRefine}($n_i.\text{content}, \text{errorMsg}, DSL$)\; \label{line:refine_format}
            $(v, \text{errorMsg}) \gets$ \texttt{SyntaxChecker}($F_i$)\; \label{line:syntax_check_again}
        }
        $n_i.F \gets F_i$\; \label{line:save_format}
    }
    $F_{\text{complete}} \gets$ \FMerge{$\mathcal{T}_{\text{DocTree}}$}\; \label{line:merge_formats}
    \Return{$F_{\text{complete}}$}\;
}

\Fn{\FMerge{$\mathcal{T}_{\text{DocTree}}$}}{
    \ForEach{$n_i \in \mathcal{T}_{\text{DocTree}}$ in \textbf{bottom-up} order}{ \label{line:foreach_bottomup}
        \If{$n_i$.hasChildren()}{ \label{line:if_children}
            $C_{\text{children}} \gets \{(n_c.\text{summary}, n_c.F) \mid \text{for each child } n_c \in \text{children of } n_i\}$\; \label{line:collect_pairs}
            $n_i.F, n_i.\text{summary} \gets $ \texttt{LLMMergeFormats}($n_i.\text{content}, n_i.F, C_{\text{children}}$)\; \label{line:merge_node}
        }
    }
    \Return{$\mathcal{T}_{\text{DocTree}}.{\text{root}}.F$}\; \label{line:return_root_format}
}
\end{algorithm}

\end{minipage}%
\hfill
\begin{minipage}{0.38\textwidth} 
    \centering
    \includegraphics[width=\linewidth]{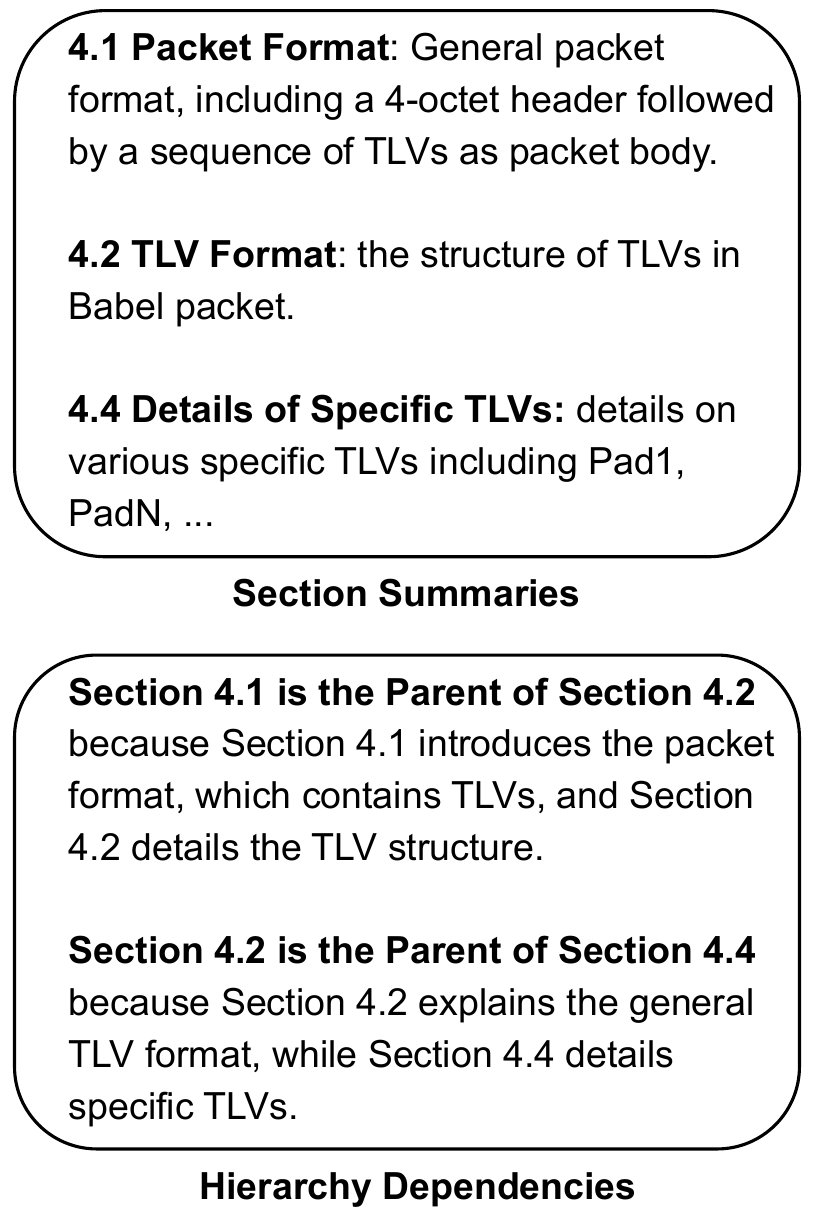}
    \caption{LLM Output for Generating Section Summaries and Identifying Hierarchy Dependences}
    \label{fig:output}
\end{minipage}
\end{figure*}

\begin{example}
For the RFC example in \Cref{fig:motivation_process} \whitecircle{A}, document sections \textit{``4.1\text{''}}, \textit{``4.2\text{''}}, and \textit{``4.4\text{''}} are listed at the same hierarchy level in the table of contents. 
\textit{``section 4.1\text{''}} introduces the general packet format, \textit{``section 4.2\text{''}}  details the \texttt{TLV} format (a component of the format specified by \textit{``section 4.1\text{''}}), and \textit{``section 4.4\text{''}}  elaborates on specific \texttt{TLV}s. 
Therefore, \textit{``section 4.2\text{''}}  should be a subcomponent of  \textit{``section 4.1\text{''}}, and \textit{``4.4\text{''}} should expand on \textit{``4.2\text{''}}. The LLM outputs for section summaries and hierarchy dependencies are listed below in \Cref{fig:output}.
After parsing the LLM's responses, \sysname adjusts the DocTree by setting the node representing ``\textit{section 4.1}\text{''}  as the parent node for ``\textit{section 4.2}\text{''}, and ``\textit{section 4.2}\text{''} as the parent node for ``\textit{section 4.4}\text{''}. 
By repeating this process across all sections, \sysname constructs the DocTree without format.
\end{example}

\vspace{-3mm}

\vspace{-2mm}

\subsubsection{Protocol Format Extraction via Divide-and-Conquer}\label{subsec:formatgen}

Algorithm \ref{alg:protocol_extraction} outlines the overall process of protocol format generation, with inputs DocTree ($\mathcal{T}_{\text{DocTree}}$) and protocol format syntax ($DSL$). Here, $DSL$ is a detailed description for format syntax defined in \Cref{syntax}, guiding LLMs to generate outputs satisfying this syntax. 

\vspace{1mm}
\noindent\textbf{Node-level Protocol Format Generation (Divide).}
The protocol format for each DocTree node is generated as described by lines 3--8 in Algorithm~\ref{alg:protocol_extraction}.
First, the content of each node and $DSL$ are passed to the LLM to produce an initial protocol format (line 3). A syntax checker then validates the format (line 4). If syntax errors are found, they are returned to the LLM for refinement (lines 5--7). This cycle of validation and refinement repeats until the format is syntax-correct. Once validated, the final format is saved in the corresponding DocTree node (line 8).
This process is repeated for every node in the DocTree (lines 2--8).

\noindent\textbf{Merge Format (Conquer).}
After generating formats for all nodes, the algorithm merges them into a single protocol format (lines 11--16). 
The merging process starts from the leaf nodes and moves upwards, following the hierarchy defined by the DocTree (line 12). 
 For each node with child nodes, the algorithm gathers the summaries and formats of its children as pairs (line 14). These pairs, represented by $C_{Children}$ (line 14), along with the content and current format of the node itself, are provided to the LLM to produce a unified format and an updated summary for that node (line 15). 
Nodes without children retain their original format and summary. This merging process continues until it reaches the root node, whose final format represents the protocol format for the entire document, resulting in a complete DocTree.
The prompt used for merging formats is as follows:

\begin{tcolorbox}[colback=gray!10, colframe=gray!60, sharp corners, title=Prompt to merge formats (LLMGenerateFormats), 
fonttitle=\scriptsize, 
boxsep=0.5mm, 
top=0.5mm, 
bottom=0.5mm] 
\scriptsize 
Task: Merge multiple protocol formats into a single comprehensive format.
Current section: \{section\};
Current format: \{format\};
Summaries and formats of child nodes: \{children\}.
\end{tcolorbox}

\begin{example}
In the DocTree shown in \Cref{fig:motivation_process} \whitecircle{B}, ``\textit{section 4.4}\text{''} serves as the parent to ``\textit{section 4.4.1}\text{''} through ``\textit{section 4.4.7}\text{''}, each defining a specific \texttt{TLV} (Type-Length-Value) format. 
Each \texttt{TLV} begins with a \texttt{Type} field, followed by its unique structure.
Since the original format for ``\textit{section 4.4}\text{''}  is empty, merging its child formats creates a unified \texttt{TLV} structure, shown in \Cref{fig:FormatGraph} (a). 
The \texttt{TLV} struct contains a \texttt{Type} field and a \texttt{Payload} field whose structure depends on the value of \texttt{Type}.
By applying this merging strategy iteratively, we obtain a complete DocTree shown in \Cref{fig:motivation_process} \whitecircle{C}, with the root node containing the complete protocol format.
\end{example}

\subsubsection{Traceability between RFC Documents and Extracted Protocol Formats}\label{subsec:trace}
In this phase, traceability is established by linking each part of the generated protocol format to its corresponding section in the original RFC document. 
The DocTree structure maintains these links by storing both the section content and its generated format in each node.
Even when multiple formats are combined into a single structure (like a \textsf{Case Type} switch), this traceability remains intact. For instance, the \texttt{RouterIdTLV} format can be traced directly back to ``\textit{section 4.4.7}\text{''}.
\subsection{Phase 2: Fine-Grained Testing by Property-Level Mutation}\label{section: stage2}
As shown in \Cref{fig:pipeline}, Phase 2 transforms the complete protocol format into a Format Graph, where each path represents a valid protocol format.
By iterating over each path in the Format Graph, both positive and negative test cases are generated. These test cases are then used to validate the parser executables. In what follows, we first describe how the format is transformed into a Format Graph in \Cref{subsec:Format Graph}, and then illustrate how test inputs are generated in \Cref{sub:test}.

\subsubsection{Format Graph Construction}\label{subsec:Format Graph}
A Format Graph is a Directed Acyclic Graph (DAG) that represents the protocol format.
Each \textsf{Primitive Type} (defined in \Cref{syntax}) is represented by a node in the Format Graph.
Complex types such as \textsf{Struct Type} and \textsf{Array Type} are
represented as subgraphs, which are connected to form a complete Format Graph.

\begin{definition}
A Format Graph is a tuple $G = (N, S, E)$, where:
\begin{itemize}[leftmargin=0.6cm]
    \item $N$ is the set of nodes, with each node $n_i$ as a tuple $n_i = (\texttt{name}, \texttt{type}, \texttt{constraint})$, representing a field with a \textsf{Primitive Type}. \texttt{constraint} defines the field-level constraint.
    \item \( S \) is the set of subgraphs, where each subgraph \( s_i = (N_i, S_i, E_i) \) is a Format Graph and represents a subgraph of the complete Format Graph.
    Subgraphs are used to represent complex types such as \textsf{Struct Type}, \textsf{Array Type}, and \textsf{Case Type} in protocol formats.
   \item \( E \subseteq (N \cup S) \times (N \cup S) \) is the set of directed edges, allowing connections between nodes, subgraphs, or both.
   Each edge is a triplet \( (x, \texttt{constraint}, y) \), indicating that \( y \) can only follow \( x \) if certain conditions, defined by \texttt{constraint}, are met. If no such condition exists, \texttt{constraint} is set to \texttt{None}. 
   This \texttt{constraint} is a structural constraint that applies primarily to \textsf{Array Type} (to define variable lengths within arrays) and \textsf{Case Type} (to map specific case values to corresponding case formats). It is distinct from the field-level constraints in each node, which apply directly to individual fields. 
\end{itemize}
Each complete path in the Format Graph is an ordered sequence of fields and constraints, representing a valid protocol format.
The combinations of these paths define all valid protocol formats.
\end{definition}

\begin{algorithm}[H]
\footnotesize
\caption{\footnotesize Format Graph Generation}
\label{alg:generate_Format Graph}
\KwIn{$F_{\text{complete}}$: Complete Protocol Format}
\KwOut{$G = (N, S, E)$: Format Graph}

\SetKwFunction{FMain}{GenerateFormatGraph}
\SetKwFunction{FSubgraph}{GenerateSubgraph}
\SetKwFunction{FStruct}{GenerateStructGraph}
\SetKwFunction{FArray}{GenerateArrayGraph}
\SetKwFunction{FCase}{GenerateCaseGraph}

\SetKwProg{Fn}{Function}{:}{}

\Fn{\FMain{$F_{\text{complete}}$}}{   
    \Return{\FStruct{$F_{\text{complete}}.\texttt{EntryStruct}$}}
}

\Fn{\FStruct{$\texttt{struct}$}}{
    $N \gets \emptyset, S \gets \emptyset, E \gets \emptyset, \texttt{prevnodes} \gets \emptyset$\;
    
    \ForEach{$\texttt{field} \in \texttt{struct.fields}$}{
        
        \uIf{\texttt{field} is PrimitiveType}{
             $n \gets (\texttt{field.name}, \texttt{field.type}, \texttt{field.constraint})$\;
             
             Connect(\texttt{prevnodes}, $n$), update $N$, $E$\;

             $\texttt{prevnodes} \gets \{n\}$\;
        }
        
        \Else{
            $s \gets \FSubgraph{\texttt{field}}$\;
            
            Connect(\texttt{prevnodes}, $s$’s start node), update $S$, $E$\;
            
           $\texttt{prevnodes} \gets \{s$'s end nodes\}\;
        }
        
    }
    \Return{$(N, S, E)$}\;
}

\Fn{\FSubgraph{$\texttt{field}$}}{
    \lIf{\texttt{field} is StructType}{\Return{\FStruct{$\texttt{field}$}}}
    \ElseIf{\texttt{field} is ArrayType}{
    \If{array length is fixed} {
        \Return{a single subgraph with sequential instances of $\texttt{field.type}$\;}
    }
    \Else{\Return{multiple subgraphs for different possible lengths\;}}
    }

    \ElseIf{\texttt{field} is CaseType}{
        \Return{a graph where each case has a subgraph, connected to the control field based on its value\;}
    }
}
\end{algorithm}

\vspace{2mm}
Algorithm \ref{alg:generate_Format Graph} constructs the Format Graph from the complete protocol format $F_{\text{complete}}$ generated in Phase 1. It begins by calling \texttt{GenerateStructGraph} on the entry struct of $F_{\text{complete}}$ (lines 1--2).
\texttt{GenerateStructGraph} constructs the graph for a \textsf{Struct Type} by first initializing empty sets for nodes \( N \), subgraphs \( S \), and edges \( E \) (line 4).
It then iterates through each field in the struct and builds the graph incrementally (lines 5--13):
(1) \textbf{Primitive Type}: Each field of Primitive Type is represented as an individual node (lines 6--9). \texttt{Connect} adds edges between the current node and all previously processed nodes (line 8), ensuring connectivity within the Format Graph.
(2) \textbf{Complex types (Struct, Array, or Case types)}: subgraphs are created recursively for each field (lines 11--13).
For \textsf{Struct Type}, a subgraph is constructed by recursively calling
\texttt{GenerateStructGraph}(line 16).
For \textsf{Array Type}, if the array length is fixed, a single sequential subgraph is created (lines 18--19). If the length is dynamic (i.e., dependent on other fields), multiple subgraphs are generated to represent different possible lengths (lines 20--21). These subgraphs are connected with the previous nodes via edges that enforce length constraints.
For \textsf{Case Type}, each case is represented by a subgraph connected to the node representing the control field (lines 22--23).
The edge between the control field node and each case subgraph enforces the constraint that the control field's value must match the corresponding case value.
By following the field order in the protocol format, the final Format Graph accurately represents the protocol structure and is ready for test case generation (line 14). 

\begin{figure}[t]
    \centering
    \includegraphics[width=\linewidth]{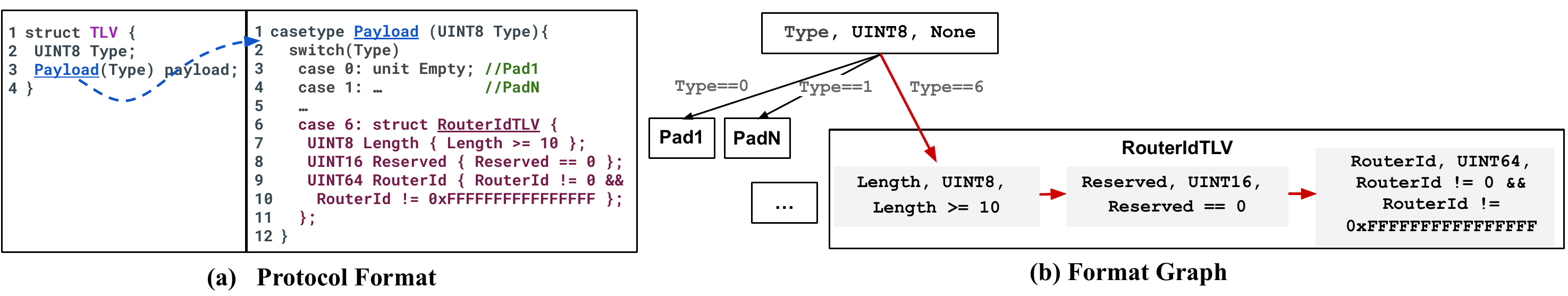}
    \caption{Transform Protocol Format to Format Graph for Test Generation}
    \label{fig:FormatGraph}
\end{figure}

\begin{example}
The format in \Cref{fig:FormatGraph} (a) is transformed into the Format Graph shown in \Cref{fig:FormatGraph} (b).
First, the algorithm creates a node for the \texttt{Type} field in the \texttt{TLV} struct, including its type (\texttt{UINT8}) and constraint (\texttt{None}).
Since the \texttt{Payload} field is a \textsf{Case Type}, it then generates a subgraph for each possible value of the control field \texttt{Type}.
For example, when \texttt{Type} equals 0, an empty subgraph representing \texttt{Pad1} is created, as \texttt{Pad1} has no additional fields. 
This subgraph is connected to the \texttt{Type} node with the constraint $\texttt{Type} == 0$. 
Similarly, when \texttt{Type} is 6, a subgraph for \texttt{RouterIdTLV} is created, containing sequential nodes for \texttt{length}, \texttt{reserved}, and \texttt{RouterId}. This subgraph is connected to the \texttt{Type} node with the constraint $\texttt{Type} == 6$.
\end{example}

\subsubsection{Fine-Grained Test Generation}\label{sub:test}
To thoroughly validate protocol parsers, we first encode each path in the Format Graph into a Z3 formula. 
Nodes (representing fields) and edges (representing additional constraints) along the path are converted into Z3 expressions, which are combined into an entire formula representing a valid format.
Then, for each path formula, both positive and negative inputs are generated:

\vspace{0.5mm}
\noindent\textbf{Positive Inputs.}
For each path formula, the Z3 SMT solver generates a satisfiable assignment, providing concrete values for the fields.
These values are used to construct a binary packet, which is tested against the target parser.
If the parser accepts the packet, it is considered temporarily consistent. If the parser rejects it, an inconsistency between the extracted format and the parser implementation is detected.
\change{Typically, protocol parsers indicate whether a packet is successfully parsed or rejected using status codes (e.g., returning 0 for success and -1 for failure) or error messages logged through specific functions. Otherwise, we annotate the expected endpoint of successful parsing. If the parser reaches this point, the packet is considered successfully parsed; otherwise, an early exit indicates rejection.}

\vspace{0.5mm}
\noindent\textbf{Negative Inputs.}
Negative inputs are generated by \textbf{property-level mutations} on positive inputs.
\textbf{Format properties} can be divided into two categories: field-level properties, which define constraints on individual fields, and structure-level properties, which address the overall format and structure of the packet.
Hence, we apply two types of mutations:
(1) Field-level mutation: This mutation negates one field value at a time against its constraint, keeping other field values unchanged. Constraints on edges (from \textsf{Case Type} or \textsf{Array Type}) are not negated in this case, as this would lead to a different path in the Format Graph.
(2) Structural mutation: This mutation adds or removes specific fields or bytes to violate the structural properties of the packet (e.g., adding or removing bytes to violate the length constraint in \textsf{Array Type}, or adding/removing a field from the packet).
Mutations are applied by negating the constraint corresponding to the selected property while adding additional constraints to enforce that all other field values remain unchanged. The modified Z3 formula is then solved again using the Z3 SMT solver to generate a mutated packet.
If the target parser rejects the mutated packet as expected, the extracted format aligns with the parser implementation. 
Otherwise, if the parser accepts the mutated packet, this reveals an inconsistency between the LLM-extracted format and the parser implementation.
\begin{example}
\label{eg:format}
For the path highlighted in red in \Cref{fig:FormatGraph} (b), the Z3 formula is: 
\footnotesize
\[
\begin{aligned}
    &0 \leq \texttt{Type} \leq 2^8-1 \land  \texttt{Type} = 6 \land  0 \leq \texttt{Length} \leq 2^8-1 \land  \texttt{Length} \geq 10 \land  0 \leq \texttt{Reserved} \leq 2^{16}-1 \\ & 
    \land \texttt{Reserved} = 0 \land  0 \leq \texttt{RouterId} \leq 2^{64}-1 \land \texttt{RouterId}\neq 0 \land \texttt{RouterId}\neq 0xFFFFFFFFFFFFFFFF
\end{aligned}
\]
\normalsize
In this formula, \texttt{Type} is constrained between $0$ and $2^8-1$ because it is of type \texttt{UINT8}. Similar for other variables.
Solving this Z3 formula produces a positive input:
\(\texttt{Type} = 6\), \(\texttt{Length} = 10\), \(\texttt{Reserved} = 0\), \(\texttt{RouterId} = 1\).
The parser accepts the binary packet constructed from this assignment, indicating temporary consistency.
Next, negative test cases are generated by mutating the positive test case. 
\change{When we negate the constraint \( \texttt{Length} \geq 10 \), the Z3 formula is:}
\change{
\footnotesize
\[
\begin{aligned}
    &0 \leq \texttt{Type} \leq 2^8-1 \land  \texttt{Type} = 6 \land  0 \leq \texttt{Length} \leq 2^8-1 \land  \neg(\texttt{Length} \geq 10) \land  0 \leq \texttt{Reserved} \leq 2^{16}-1 \\ & 
    \land \texttt{Reserved} = 0 \land  0 \leq \texttt{RouterId} \leq 2^{64}-1 \land \texttt{RouterId}\neq 0 \land \texttt{RouterId}\neq 0xFFFFFFFFFFFFFFFF  \\ &\land \texttt{Type} = 6 \land \texttt{Reserved} = 0 \land \texttt{RouterId} = 1
\end{aligned}
\]
\normalsize
}
\change{Solving this Z3 formula produces a negative input:}
\change{\(\texttt{Type} = 6\), \(\texttt{Length} = 0\), \(\texttt{Reserved} = 0\), \(\texttt{RouterId} = 1\). This negative input changed the value of \texttt{Length} to 0, while keeping other field values unchanged.}
The parser correctly rejects it, confirming its adherence to the specification on this field-level property.
However, when we set \texttt{RouterId} to 0 — violating the format constraint: $\texttt{RouterId}\neq 0$ — the parser still accepts the packet. This indicates an inconsistency, as the parser does not enforce this constraint on \texttt{RouterId} required by the LLM-extracted protocol format. 
\end{example}

\subsection{Phase 3: Traceability-Assisted Inconsistency Identification}\label{section: stage3}
As shown in \Cref{fig:pipeline}, Phase 3 identifies each inconsistency. 
When an inconsistency is detected during testing, there are two potential causes: an error in the LLM-extracted protocol format (as the input is generated based on this format) or a bug in the implementation. Therefore, inconsistencies should not be immediately treated as implementation bugs. 
An additional validation step is needed to identify the source of the inconsistency by cross-referencing relevant sections of the RFC document.
To achieve this, we backtrace the inconsistency to its corresponding RFC section in the document, retrieve that section, and prompt the LLM to diagnose it with the following prompt:

\begin{tcolorbox}[colback=gray!10, colframe=gray!60, sharp corners, title=Prompt to identify inconsistencies, 
fonttitle=\scriptsize, 
boxsep=0.5mm, 
top=0.5mm, 
bottom=0.5mm] 
\scriptsize 
Task: \{Constraint\} is allowed by [myformat/parser] but not by [myformat/parser]. According to the RFC section: \{Section\}, identify whether myformat or parser is correct, and provide evidence from the RFC section.
\end{tcolorbox}

This process helps determine whether the inconsistencies are caused by mistakes in the extracted format or bugs in the implementation.
If the LLM validator identifies an inconsistency as an implementation error, we report it as a bug. 
On the other hand, if it is identified as a format extraction mistake, we use this feedback to further refine the extracted format.

\vspace{1mm}
\noindent\textbf{Traceability between Inconsistencies and RFC Documents.}
Each inconsistency is directly tied to a specific constraint or set of constraints.
Since traceability between the format and the corresponding RFC section is maintained, any inconsistency can be traced back to the exact section of the RFC where the relevant constraint is defined.
This traceability-based approach ensures that every step, from extracting the format to finding bugs, is linked to the original RFC document. 

\begin{example}
Consider the inconsistency regarding \texttt{RouterId} introduced in \Cref{eg:format}.
This constraint comes from the subgraph \texttt{RouterId} shown in \Cref{fig:FormatGraph} (b), which is derived from \texttt{RouterIdTLV} in the LLM-extracted protocol format shown in \Cref{fig:FormatGraph} (a).
\texttt{RouterIdTLV} corresponds to RFC document ``\textit{section 4.4.7}\text{''} (\Cref{fig:motivation_process} \whitecircle{B}). 
So we backtrace and retrieve the content of RFC ``\textit{section 4.4.7}\text{''} and provide it to the LLM validator to identify the root cause of the inconsistency. 
The LLM validator determines that it is a bug in the implementation instead of the extracted format. 
So we report the bug to the developers. It is confirmed and now fixed as shown in  \Cref{fig:motivation_code}.
\end{example}

\section{Evaluation}\label{sec:evaluation}
We implement our tool using OpenAI's GPT-4o API~\cite{gpt4o} and the Z3 SMT solver (version 4.11.2)~\cite{z3}. 
\change{We choose GPT-4o for its strong natural language understanding and DSL grammar extraction capabilities ~\cite{m2024natural}, which are essential for interpreting protocol specifications from RFC documents.
Additionally, GPT-4o has shown strong performance in software testing~\cite{Software-System-Testing, 10851734}, including protocol validation~\cite{yang2025chathttpfuzz}.}
We set the temperature to 0 for minimal randomness and better reproducibility.
For protocol format extraction (\Cref{section: stage1}), we use everparse~\cite{syntaxchecker} as the syntax checker.
To evaluate the effectiveness of our approach, we conduct experiments to address the following research questions.
\begin{itemize}[leftmargin=0.6cm]
    \item \textbf{RQ1}: How accurate are the message formats extracted from the RFC documents?
    \item \textbf{RQ2}: How effective is \sysname in inconsistency identification and bug detection?
    \item \textbf{RQ3}: How effective is \sysname compared to existing approaches?
    \item \textbf{RQ4}: How effective is each component of \sysname?
\end{itemize}

\subsection{Dataset}
To evaluate \sysname's capability to support multiple programming languages, we construct a new protocol dataset, as existing datasets~\cite{Pardiff, chatafl} only include protocol implementations in C or C++. 
We filter GitHub repositories related to network protocol implementations with over 2,000 stars and actively maintained. To increase diversity in implementations, we select three repositories, each using a different programming language: C, Go, and Python. Based on these criteria, we choose FRR~\cite{frr}, Go Networking~\cite{gonet}, and Impacket~\cite{impacket}.
For each repository, we select three protocols with RFC documents longer than 40 pages.
In total, we get nine different protocols with a maximum document length of 104 pages and an average of 62 pages per protocol.
For each protocol, we download the corresponding RFC documents from the IETF DataTracker~\cite{DataTracker} as input for \sysname.
For each protocol implementation, we built the parsing executables, as our tool does not require access to source code.
This makes our method applicable to both source-available and source-unavailable scenarios.
The protocol and codebase information is available in Table~\ref{tab:bench}.

\begin{table}[t]
    \centering
    \scriptsize
    \caption{Protocol Dataset and Ground Truth Format.}
    \label{tab:bench}

    \resizebox{\columnwidth}{!}{
    \begin{tabular}{l||lllp{42mm} | cccccc}
        \toprule
         \multirow{2}{*}{\textbf{Protocol}} &  \multicolumn{4}{c|}{\textbf{Dataset}} &  \multicolumn{6}{c}{\textbf{Ground Truth Format}} \\
         [0.5ex] 
 \cmidrule(lr){2-5} \cmidrule(lr){6-11} 
         &\multirow{1}{*}{\textbf{RFC (Pages)}}  & \multirow{1}{*}{\textbf{Repo.}} &\multirow{1}{*}{\textbf{Lang.}} & \multirow{1}{*}{\textbf{Description}}  & \multirow{1}{*}{\textbf{Msg Type}} &\multirow{1}{*}{\textbf{Field}} & \multirow{1}{*}{\textbf{Indep. Constr.}}& \multirow{1}{*}{\textbf{Dep. Constr.}} &\multirow{1}{*}{\textbf{LoC}} &\multirow{1}{*}{\change{\textbf{Time}}}  \\
         
         \midrule
          BABEL & 8966 (54) & FRR & C &Distance-vector routing protocol&11 & 97 & 66 & 8 & 171 & \change{5h}\\
          BFD   & 5880 (49) & FRR & C&Bidirectional forwarding detection& 6 & 43 & 22 & 2 & 69 & \change{2h} \\
          BGP-4 & 4271 (104)& FRR & C&Border Gateway Protocol 4 &4 & 62 & 46 & 6 & 164 & \change{6h}\\
          IPv4  & 791 (45) & Go Net & Go& Internet protocol v4 & 1 & 14 & 3  & 1 & 16 & \change{0.5h} \\
          ICMPv4 & 791, 792 (66)& Go Net &Go& Internet Control Message Protocol for IPv4 &11 & 73 & 27 & 3 & 106 & \change{2h} \\      
          ICMPv6 & 8200, 4443 (66) & Go Net &Go& Internet Control Message Protocol for IPv6 & 6 & 29 & 14 & 0 & 66 & \change{2h} \\    
          IPv6& 8200 (42) & Impacket &Python&Internet protocol v6 & 1 & 8   & 1  & 0 & 11 & \change{0.5h} \\
          DHCP & 2131 (45) & Impacket  &Python& Dynamic host configuration protocol & 1 & 16 & 4  & 1 & 21 & \change{0.5h} \\
          TCP & 793 (85) & Impacket  &Python& Extensible Authentication Protocol & 1 & 23 & 6  & 4 & 43 & \change{2h} \\
         \bottomrule
    \end{tabular}
    }
\end{table}

\subsection{RQ1: Effectiveness of \sysname on Message Format Extraction}
\label{sec:rq1}

\subsubsection{Setup and Metrics.}
To evaluate the accuracy of protocol formats generated by \sysname, we first establish a ground truth for network protocol formats. Two authors, each with more than three years of expertise in network protocols, independently reviewed the relevant RFC documents and manually wrote the input formats. 
They tracked the time taken to complete the formats for each protocol, compared their results, discussed inconsistencies, and reached a consensus.

To quantify the correctness of extracted formats, we define format metrics across five element types: Message Types, Field Names/Types, Independent Constraints, and Dependent Constraints, as introduced in \Cref{syntax}.  Table~\ref{tab:bench} lists the count of each element type for each protocol's ground truth format (column \textit{``Ground Truth Format\text{''}}), the line count for each (column \textit{``LoC\text{''}}), \change{and the average manual time spent labeling the ground truth (column \textit{``Time\text{''}})}.  Column \textit{``Field\text{''}} in column \textit{``Ground Truth Format\text{''}} represents the count for Field Names and Field Types. Since the number of field names and types is identical, we merge them into a single column. The statistics in the table reflect the complexity of protocol formats. For example, BABEL has 11 message types, 97 fields, and 74 constraints (66 independent, 8 dependent), totaling 171 LoC, and takes an average of 5 hours to manually construct the ground truth format, indicating high complexity.
Simpler protocols like IPv4 and IPv6 have fewer fields and constraints, and require less time to manually write the ground truth formats. 
We then compare \sysname’s extracted formats with ground truth, measuring precision  (the proportion of correct extracted elements) and recall (the proportion of ground truth elements accurately captured) for each element type.
For example, for Message Types, the precision and recall are calculated as follows:

\scriptsize

\[
\text{Precision} = \frac{\text{Correct Message Types in Extracted Format}}{\text{Total Message Types in Extracted Format}}, \quad
\text{Recall} = \frac{\text{Ground Truth Message Types Covered by Extracted Format}}{\text{Total Message Types in Ground Truth Format}}.
\]

\normalsize
Since the extracted element count or ground truth element count can be zero for each element type, making precision or recall undefined, we use ``-\text{''} to indicate these cases. 
\change{Comparing the ground truth with the formats extracted by \sysname takes an estimated fifteen minutes per protocol.}

\subsubsection{Results}

\begin{table*}[t]
\centering
\scriptsize
 \caption{\sysname Protocol Format Extraction Results: Precision/Recall(\%). 
 }

 \label{tab:ourtool}
 \begin{tabular}{l || ccccc} 
 \toprule
 \scriptsize
 \multirow{1}*{\textbf{Protocol}} 
& \textbf{Msg Type}& \textbf{Field Name} & \textbf{Field Type} & \textbf{Indep. Constr.} & \textbf{Dep. Constr.}\\ 
\midrule
 BABEL  & 100/100 & 100/91  &95/87 & 98/86 &100/75\\
 BFD    & 100/100 & 100/100 & 100/100 & 100/100 & -/0\\
 BGP-4  & 100/100 &  95/95 & 94/94 & 98/63 & 100/50\\
 IPv4   & 100/100 & 100/100&100/100 & 100/33 & 100/100\\
 ICMPv4 & 100/100 & 100/100&  100/93 & 100/96 & 100/33\\
 ICMPv6 & 100/100 & 94/100 & 90/97 & 100/100 &  -/-\\
 IPv6   & 100/100 & 100/100 & 100/100& 100/100 & -/-\\
 DHCP   & 100/100 & 100/100 & 94/94 & 83/75 & 0/0\\
 TCP    & 100/100 & 100/78  & 83/65 & 67/33 & 0/0 \\
 \midrule
 Total & 100/100& 99/95 & 94/91 & 98/82 & 73/44\\
\bottomrule
\end{tabular}
\end{table*}

The results are shown in \Cref{tab:ourtool}. \sysname achieves 100\% precision and recall in extracting message types across all protocols. Field names and field types are also accurately identified, achieving overall precision and recall of 99\%/95\% for field names and 94\%/91\% for field types.
Independent constraints are well-detected across the nine protocols, with a precision of 98\%. In terms of recall, the tool identified 155 out of 189 ground truth independent constraints, resulting in an average recall of 82\%. 
Dependent constraints are particularly challenging,  as they require understanding how different fields influence one another, often through implicit rules or cross-references within the protocol specification. These complexities make accurate extraction difficult for LLMs. Overall, \sysname achieves a high precision of 73\%.
For the ICMPv6 and IPv6 protocols, since there is no dependent constraint in either the ground truth (\Cref{tab:bench}) or the LLM-extracted format, the precision and recall are both denoted with ``-\text{''}. For BFD, since the LLM-extracted format contains no dependent constraints,  its precision is denoted with ``-\text{''}.
\change{Among these protocols, TCP shows a drop in performance due to the handling of the \texttt{DataOffset} field, which involves header alignment and bit-level offsets. These implicit constraints are not straightforward to extract from natural language descriptions, leading to reduced accuracy when interpreting the precise field layout in the TCP header. Further discussion can be found in Section~\ref{discuss}. }

\noindent\textbf{Conclusion.} \sysname excels in extracting protocol formats from RFC documents, achieving over 90\% precision and recall for most elements, with high precision on challenging element types.

\subsection{RQ2: Effectiveness of Inconsistency Identification and Bug Detection}\label{RQ3}

\subsubsection{Setup and Metrics.}
We evaluate \sysname on both inconsistency identification and bug detection. We create one positive test case per format path, which already achieves great performance. 
Generating multiple cases per path could enhance bug detection, but would significantly increase the manual effort required to verify inconsistencies. Thus, we use a single representative case per path, with negative test cases determined by property counts along the path. Also, for \textsf{Array Type} with variable length, we generate test cases containing zero and one element.

\noindent\textbf{Inconsistency Identification.}
We first record the number of inconsistencies detected by \sysname in Phase 2 (\Cref{section: stage2}), including logical inconsistencies (e.g., the extracted constraint mismatches with the implementation) and crashes. 
In Phase 3 (\Cref{section: stage3}), \sysname classifies each inconsistency as either an \textit{Implementation Error} or a \textit{Format Extraction Error} (a mistake in the extracted format), with crashes always classified as \textit{Implementation Errors}.
We manually check each classification to assess the inconsistency identification accuracy of \sysname.
\change{It takes an estimated five minutes to check each inconsistency.}

\noindent\textbf{Bug Detection.}
For each inconsistency classified as an \textit{Implementation Error}, we record whether it is a logical or crash issue. We report the number of \textit{unique} and \textit{new} (i.e., previously unknown) bugs detected, as well as the \textit{confirmed} bugs.

\vspace{-2mm}
\subsubsection{Results}
The results of inconsistency identification and bug detection are in \Cref{tab:inconsistency_bug}.

\noindent\textbf{Inconsistency Identification.}
As shown in column \textit{``Detected Incons.\text{''}}, \sysname detects a total of 90 (86 logical and 4 crash) inconsistencies across all protocols. In column \textit{``Identified Incons.\text{''}}, 79 (75 logical and 4 crash) of the 90 inconsistencies are identified as \textit{Implementation Errors}, and 11 as \textit{Format Extraction Errors}.
Our manual evaluation confirms that \sysname correctly identifies 87 inconsistencies: 76 (72 logical and 4 crash) of 79 \textit{Implementation Errors} and all 11 \textit{Format Extraction Errors}, achieving an overall 97\% accuracy and 100\% accuracy for seven protocols. Since crashes are always \textit{Implementation Errors}, the identification accuracy for crashes is 100\%. Among 86 detected logical inconsistencies, 83 (72 logical \textit{Implementation Errors}, 11 \textit{Format Extraction Errors}) are accurately classified, achieving 97\% accuracy, highlighting \sysname’s effectiveness in distinguishing between implementation and format extraction errors.

\begin{table}[]
\centering
\caption{\sysname Results on Inconsistency (Incons.) Identification and Bug Detection}
\label{tab:inconsistency_bug}
\resizebox{\textwidth}{!}{%
\begin{tabular}{l|| cc|ccc|ccc|cccc|ccc}
\toprule
\multirow{3}*{\textbf{Protocol}}
& \multicolumn{2}{c|}{\textbf{Detected Incons.}} & \multicolumn{3}{c|}{\textbf{Identified Incons.}} & \multicolumn{3}{c|}{\textbf{Correctly Identified Incons.}} & \multicolumn{4}{c|}{\textbf{Identification Acc (\%)}} & \multicolumn{3}{c}{\textbf{\# Bugs}} \\ 
\cmidrule(lr){2-3} \cmidrule(lr){4-6} \cmidrule(lr){7-9} \cmidrule(lr){9-13} \cmidrule(lr){14-16}
&  &  & \multicolumn{2}{c}{\textbf{Impl. Error}} & \multirow{2}{*}{\textbf{Format Error}} & \multicolumn{2}{c}{\textbf{Impl. Error}} & \multirow{2}{*}{\textbf{Format Error}} & \multicolumn{2}{c}{\textbf{Impl. Error}} & \multirow{2}{*}{\textbf{Format Error}} & \multirow{2}{*}{\textbf{Overall}} &  &  &  \\ \cline{4-5} \cline{7-8} \cline{10-11}
 & Logical & Crash & Logical & Crash &  & Logical & Crash &  & Logical & Total &  &  & Unique & New & Confirmed \\
\midrule
BABEL & 27 & 0 & 23 & 0 & 4 & 23 & 0 & 4 & 100 & 100 & 100 & 100 & 23 & 23 & \change{23} \\
BFD & 12 & 0 & 10 & 0 & 2 & 10 & 0 & 2 & 100 & 100 & 100 & 100 & 10 & 10 & 10 \\
BGP-4 & 3 & 0 & 0 & 0 & 3 & 0 & 0 & 3 & 100 & 100 & 100 & 100 & 0 & 0 & 0 \\
IPv4 & 2 & 0 & 2 & 0 & 0 & 2 & 0 & 0 & 100 & 100 & 100 & 100 & 2 & 2 & 0 \\
ICMPv4 & 16 & 0 & 16 & 0 & 0 & 16 & 0 & 0 & 100 & 100 & 100 & 100 & 16 & 16 & 0 \\
ICMPv6 & 10 & 0 & 8 & 0 & 2 & 8 & 0 & 2 & 100 & 100 & 100 & 100 & 8 & 8 & 0 \\
IPv6 & 1 & 2 & 1 & 2 & 0 & 1 & 2 & 0 & 100 & 100 & 100 & 100 & 2 & 1 & 1 \\
DHCP & 4 & 2 & 4 & 2 & 0 & 3 & 2 & 0 & 75 & 100 & 100 & 83 & 5 & 5 & 2 \\
TCP & 11 & 0 & 11 & 0 & 0 & 9 & 0 & 0 & 82 & 100 & 100 & 82 & 3 & 3 & 0 \\
\midrule
Total & 86 & 4 & 75 & 4 & 11 & 72 & 4 & 11 & 97 & 100 & 100 & 97 & 69 & 68 & \change{36} \\
\bottomrule
\end{tabular}%
}
\end{table}

\noindent\textbf{Bug Detection.}
As shown in column \textit{``Correctly Identified Incons.\text{''}}, 72 logical and 4 crash issues detected by \sysname are true bugs, including 69 unique bugs. This demonstrates \sysname's ability to detect diverse bugs through fine-grained test generation, which thoroughly covers the input space and behaviors. Notably, 68 of these bugs are new, underscoring \sysname's effectiveness. We reported them to the developers, with 36 confirmed so far and others pending review. \sysname generated a PoC (Proof of Concept) for each bug.
Of the 36 confirmed bugs, 17 are fixed and merged, 18 have approved pull requests pending merge, and 1 remains unresolved.

\noindent\textbf{Conclusion.} \sysname demonstrates strong effectiveness in inconsistency identification and bug detection, achieving 97\% identification accuracy. It successfully detects 68 new bugs, with 36 confirmed, across nine protocols. Each bug is generated with a PoC.

\subsection{RQ3: Comparative \sysname with Baseline Methods}
\label{subsec:baseline}
\subsubsection{Setup and Metrics.}
We compare \sysname with two baseline methods to further evaluate \sysname's format extraction performance and bug detection effectiveness.

\noindent\textbf{ChatAFL~\cite{chatafl}}. We compare \sysname's input format extraction capability with a state-of-the-art LLM-based testing tool, ChatAFL, which also uses LLMs to retrieve protocol formats. To ensure a fair comparison, all experiments involving LLMs use GPT-4o as our approach.

\noindent\textbf{ParDiff~\cite{Pardiff}.} We compare \sysname to ParDiff, a state-of-the-art differential testing tool to detect network protocol parsing bugs. 
ParDiff, built on LLVM, requires access to source code and only supports parsers written in C.
In contrast, \sysname does not require source code and can work directly with parser executables. 
Although this makes the comparison unfair, we still compare bug detection results by running ParDiff on the BABEL, BFD, and BGP-4 protocols, as it does not support the Python and Go implementations used in six other protocols. We do not compare bug detection with ChatAFL, as ChatAFL only detects crashes.

\subsubsection{Results}
\Cref{tab:baseline} presents the protocol format extraction comparison between \sysname and ChatAFL. 
ChatAFL struggles to capture accurate protocol formats, partly because it lacks support for specifying individual field types (column \textit{``Field Type\text{''}}), a crucial feature in binary protocol formats. Additionally, ChatAFL struggles with complex constraints (e.g., dependencies) and is limited to expressing basic constraints, such as concrete values. These limitations greatly reduce its effectiveness in accurately extracting network protocol formats.

\begin{table*}[t]
\begin{minipage}[t]{0.58\textwidth} 
    \centering
    \caption{\sysname vs. Baseline on Protocol Format Extraction: Precision/Recall (\%).}
\label{tab:baseline}
    \resizebox{\textwidth}{!}{ 
        \begin{tabular}{l||ccccc}

            \toprule
            Approach & Msg Type & Field Name & Field Type & Indep. Constr. & Dep. Constr. \\
            \midrule
            \sysname &  \textbf{100}/\textbf{100}& \textbf{99}/\textbf{95} & \textbf{94}/\textbf{91} & \textbf{98}/\textbf{82} & \textbf{73}/\textbf{44}\\
            ChatAFL~\cite{chatafl} &  89/55 & 78/35 &-/- & 81/10 & -/0 \\
            \bottomrule
            \end{tabular}
    }
\end{minipage}%
\hfill 
\begin{minipage}[t]{0.4\textwidth} 
    \centering
    \scriptsize
    \caption{\sysname vs. Baseline on Bug Detection for BABEL, BFD, and BGP-4.
    }
\label{tab:baseline_bug}
    \resizebox{0.88\textwidth}{!}{ 
        \begin{tabular}{c||ccc}

            \toprule
            Approach & Unique & New & Confirmed \\ \midrule
            \sysname & \textbf{33} & \textbf{33} & \textbf{33}\\
            ParDiff~\cite{Pardiff} & 4 & 1& 1 \\
            \bottomrule
        \end{tabular}
    }
\end{minipage}
\end{table*}

\Cref{tab:baseline_bug} shows the comparison bug detection results  of \sysname and ParDiff. \sysname detects 33 unique bugs in the FRR project, all of which are new. In contrast, ParDiff detects only 4 unique bugs in the FRR project, including only 1 new bug. This is due to two main factors: first, many bugs are shared across both tested implementations, which ParDiff cannot detect through differential analysis. Second, ParDiff halts bisimulation on each FSM path after the first state mismatch, missing bugs in further state transitions along that path.

\noindent\textbf{Conclusion.}  \sysname achieves high performance on protocol format extraction and bug detection, outperforming state-of-the-art methods.

\subsection{RQ4: Ablation Studies}
\change{To evaluate the effectiveness of each \sysname design in mitigating LLM hallucinations, we conduct ablation studies to assess the impact of traceability-assisted format refinement (\Cref{sec:ablation_refine}) and divide-and-conquer strategy (\Cref{sec:ablation_divide}) on protocol format extraction, as well as the impact of traceability on inconsistency identification (\Cref{sub:ablation_incon}) and fine-grained testing on inconsistency and bug detection (\Cref{sec:ablation_bug}).}

\begin{table}[t]
    \centering
    \scriptsize
    \caption{The precision (\%) and recall (\%) of ablations}
    
    \label{tab:ablation_extraction}
        \begin{tabular}{l||ccccc}

            \toprule
            Approach & Msg Type & Field Name & Field Type & Indep. Constr. & Dep. Constr. \\
            \midrule
            \sysname &  \textbf{100}/\textbf{100}& \textbf{99}/\textbf{95} & \textbf{94}/\textbf{91} & \textbf{98}/\textbf{82} & 73/\textbf{44}\\
            \sysname~ - refine &  \textbf{100}/\textbf{100} & \textbf{99}/94   & 93/89   & 88/68   & \textbf{85}/\textbf{44} \\
            \sysname ~ - refine - divide and conquer & 97/71   & 91/54  & 65/37 & 79/26  & 0/0\\
            \bottomrule
            \end{tabular}
\end{table}

\subsubsection{Format Extraction without Traceability-Assisted Format Refinement}
\label{sec:ablation_refine}
\change{We evaluate the impact of traceability-assisted format refinement (\Cref{section: stage3}) on the extracted protocol format, with results presented in \Cref{tab:ablation_extraction}. Specifically, we compare the protocol format after refinement (Phase 3) in row ``{\sysname}\text{''} to the format without refinement (after Phase 1) in row \textit{``\sysname~- refine\text{''}}.}

Overall, refinement improves the quality of extracted protocol formats, increasing the precision and recall of independent constraints (column \textit{``Indep. Constr.\text{''}}) by 11\% (88\% to 98\%) and 21\% (68\% to 82\%). 
\change{This demonstrates that traceability-assisted format refinement effectively mitigates LLM hallucinations by correcting inaccurately generated constraints.}
However, \sysname’s precision of dependent constraints slightly decreases after refinement. This is because, without refinement, fewer dependent constraints are generated with high precision (85\%). After refinement, the model identifies two additional missing constraints but produces inaccurate formulas due to the complexity of dependencies, leading to a minor precision drop. Despite this, the refinement effectively enhances the overall accuracy and completeness of the extracted protocol formats.

\subsubsection{Format Extraction without Divide-and-Conquer}
\label{sec:ablation_divide}
To evaluate the effectiveness of the divide-and-conquer approach (\Cref{section: stage1}), we compare the format extraction performance in Phase 1 (\Cref{tab:ablation_extraction} row \textit{``\sysname - refine\text{''}}) with the performance by directly feeding the entire RFC document into the LLM (with unnecessary lines removed to fit within the model's input window) in row \textit{``\sysname - refine - divide and conquer\text{''}}. 
The results reveal a consistent and significant performance drop across all five element types when using the full document without divide-and-conquer. For example, field name recall decreases from 94\% to 54\%, and field type recall drops from 89\% to 37\%. 
Extraction of constraints suffers even more, with independent constraint recall falling by 55\% (from 68\% to 26\%) and all dependent constraints being missed.
These findings highlight that the divide-and-conquer strategy effectively improves the accuracy of protocol format extraction.

\subsubsection{Inconsistency Identification without Traceability-Assisted Inconsistency Identification}\label{sub:ablation_incon}
\change{We conduct an ablation study to evaluate the impact of traceability-assisted inconsistency identification (\Cref{section: stage3}). Instead of using the corresponding RFC section, the LLM is provided with the entire RFC document for each inconsistency detected in Phase 2. The results are shown in \Cref{tab:ablation_bug} row \textit{``\sysname - traceability\text{''}}. Without traceability, the LLM fails to identify any \textit{Format Extraction Errors}, causing accuracy in this category to drop from 100\% to 0\%. Additionally, the overall accuracy decreases from 97\% to 84\%, underscoring the importance of traceability-assisted inconsistency identification in effectively distinguishing implementation errors from format extraction errors.}

\subsubsection{Inconsistency and Bug Detection without Fine-Grained Testing}
\label{sec:ablation_bug}
\begin{table}[t]
\centering
\scriptsize
\caption{{Impact of and Fine-Grained Testing and Traceability-Assisted Inconsistency Identification}}
\label{tab:ablation_bug}
\resizebox{\textwidth}{!}{%
\begin{tabular}{l||ccc|cc|ccc}
\toprule
\multirow{2}{*}{\textbf{Approach}} & \multicolumn{3}{c|}{\textbf{Incons. Identification Accuracy}} & \multicolumn{2}{c|}{\textbf{\# Incons.}} & \multicolumn{3}{c}{\textbf{\# Bugs}} \\
  [0.5ex] \cmidrule(lr){3-9}
 & Logical & Format Error & Total & Logical & Crash & Unique & New & Confirmed \\
 \midrule

\sysname & \textbf{97\%} & \textbf{100\%} & 97\% & 75 & 4 & \textbf{69} & 68 & 36 \\
\change{\sysname - traceability} &\change{84\%} &\change{0\%} & \change{84\%} & \change{75} & \change{4} & \change{-} & \change{-} & \change{-} \\
\sysname - fine grained testing - traceability & 4\% &0\%& 90\% & 22 & 211 & 4 & 4 & 0 \\ 
\bottomrule
\end{tabular}%
}
\end{table}

We conduct an ablation study to assess the impact of fine-grained testing. In this setup, the full format produced in Phase 1 (\Cref{section: stage1}) is encoded as a single Z3 formula, used for generating 50 positive and 100 negative test cases (both more than the amount that \sysname generates for each protocol).
These cases are then executed against the target parser to obtain parsing results.
Since inputs are generated based on the entire format, individual fields and document sections are unknown during the diagnosis step (\Cref{section: stage3}). Therefore, in this setting, during diagnosis, the LLM is provided with binary input and the complete RFC document (instead of only the relevant RFC section supported by traceability) to help locate the source of any inconsistencies. We then compare the performance of inconsistency identification accuracy and bug detection of this approach against \sysname.

The comparison results are shown in \Cref{tab:ablation_bug} row \textit{``\sysname - fine grained testing - traceability\text{''}}. Without traceability-assisted inconsistency identification, accuracy in identifying real implementation errors drops sharply from 97\% to 4\%, highlighting that, without traceability, LLMs struggle to accurately detect logical inconsistencies due to hallucinations. 
By linking inconsistencies to specific RFC sections, traceability provides essential context to mitigate hallucinations and enhance inconsistency identification accuracy.
Column \textit{``\#Incons.\text{''}} lists the number of inconsistencies detected by each tool, while \textit{``\#Bugs\text{''}} shows the number of detected bugs. Without fine-grained property-level mutation, most detected issues are crashes (211 out of 233), with only 4 unique bugs identified. This indicates that inputs generated based on the entire format fail to \textit{comprehensively} test the input space. In contrast, \sysname, using fine-grained testing, detected 75 logical inconsistencies and 4 crashes, with 69 unique bugs detected. This demonstrates that fine-grained test generation enables thorough input space exploration, triggering more logical inconsistencies and unique bugs.

\section{Threats to Validity.}\label{sec:threats}
An internal threat to validity is the manually established ground truth (\Cref{sec:rq1}). Human error or bias in defining input formats may impact evaluation accuracy. To address this, two experts independently draft the formats, compare results, and resolve discrepancies to reach a consensus~\cite{xie2023impact, 9678600}.


An external threat is the potential for bugs in the RFC documents. Our method assumes the document is of high quality and treats it as the ground truth. In fact, protocol documents are generally regarded as reliable oracles~\cite{InternetStandard, IETFRFC}. Additionally, in our experiments, we did not observe any document bugs.
Another threat is potential data leakage. Since LLMs are pretrained on vast datasets, the model may have seen the documents we tested, which could undermine the effectiveness of the divide-and-conquer approach for protocol format extraction (\Cref{section: stage1}). To mitigate this, we compare with 
a baseline where the entire document is fed directly to GPT-4o (\Cref{sec:ablation_divide}). These experiments validate the effectiveness of our divide-and-conquer design.

\section{Limitations and Future Work}
\label{discuss}
\noindent\change{\textbf{LLMs in Extracting Complex Protocol Formats.} \sysname uses LLMs (e.g., GPT-4o) to interpret RFC documents.
While LLMs generally perform well when extracting clearly stated protocol formats, they struggle with implicit details like padding in the TCP header. 
For instance, RFC 793~\cite{RFC793} specifies that "the TCP header padding is used to ensure that the TCP header ends and data begins on a 32-bit boundary", but it does not explicitly describe how padding is applied or calculated. Correctly constructing the padding field requires reasoning about alignment constraints between header size, data offset, and structure layout, which are only implicitly conveyed in the document. As a result, LLMs struggle to accurately infer these relationships, leading to hallucinations.
Future work could enhance format extraction by fine-tuning LLMs on domain-specific datasets, developing rule-based post-processing for implicit formats, and incorporating self-verifying mechanisms.}

\vspace{1mm}
\noindent\change{\textbf{\sysname in Handling Non-standardized Documentation.} \sysname extracts protocol formats from structured RFCs, as most protocols have Standards Track RFCs as official standards.
For protocols without RFC documentation, relying on non-standardized or incomplete sources is often unreliable. Future work could integrate alternative sources (e.g., technical manuals, legacy parsers) and human-in-the-loop feedback to improve adaptability.}

\section{Related Work}\label{sec:relatedwork}
\noindent
\textbf{Conventional Fuzzing.}
Conventional network protocol fuzzing (e.g., BooFuzz~\cite{boofuzz} and SAGE~\cite{Msrsage}) identifies bugs by exposing crashes.
Netlifter~\cite{shi2023lifting} combines conventional fuzzing with static analysis. It leverages static symbolic analysis to collect path constraints and generate varied test cases for fuzzing.
However, they still rely on crashes as oracles to detect bugs.

\vspace{1mm}
\noindent
\textbf{Differential Analysis.}
Differential analysis~\cite{Pardiff, KIT, ODIT} has been widely used for bug detection. Existing differential analysis finds bugs by comparing multiple implementations. 
Static differential analysis tools locate semantic differences by comparing models derived from these implementations. 
For example, ParDiff~\cite{Pardiff} compares protocol formats derived from different parser implementations to find inconsistencies. 
Dynamic differential testing techniques like DPIFuzz~\cite{reen2020dpifuzz} feed different implementations with the same input and compare their execution behaviors.
While these static and dynamic approaches can detect semantic bugs (i.e., silent violations of protocol rules), they cannot identify bugs present in both implementations, a limitation inherent to differential analysis. To address this, \sysname compares each implementation directly with its official specification, enabling the detection of such overlooked bugs.

\vspace{1mm}
\noindent
\textbf{Traditional Static Analyzers and Model Checking.} 
Static analyzers (e.g., Pistachio~\cite{Pistachio}) detect protocol-specific bugs by checking implementations against data-dependent rules, while model checking~\cite{MusuvathiE04, DiazCRP04, Magic} verifies behavior using formal models. Both methods require extensive manual effort to define rules or models, as protocols are typically described in natural language. \sysname overcomes this by using LLMs to automatically extract protocol formats from RFC documents.

\vspace{1mm}
\noindent
\textbf{LLM-based Approaches.}
LLMs have shown effectiveness across various domains in software engineering, including code generation~\cite{ding2024cycle, liu2024evaluating, zhu2024deepseek, LiangXZZD0CWF24}, 
software testing~\cite{deng2023large, kang2023large, yang2023whitefox, 10.1145/3643769}, 
program analysis~\cite{wang2024dataflow, WangZSX024, 10.1145/3649828, li2025hitchhiker, guo2025repoaudit}, comment/specification generation~\cite{wen2024enchanting, WangZW024, geng2023empirical,xie2023impact}, and automated repair~\cite{jiang2023impact, zhang2024autocoderover, llm-vul-1, llm-vul-2}.
For network protocol testing, LLM-based methods construct protocol models automatically to assist fuzzing. 
ChatAFL~\cite{chatafl} leverages LLMs' prior knowledge to generate protocol formats, but only works well on simple and well-known protocols, as it does not directly learn the format details from the document content.
mGPTFuzz~\cite{mGPTFuzz} extracts finite-state machines (FSMs) from RFCs but lacks detailed protocol formats and symbolic constraints.
To bridge these gaps, \sysname uses a divide-and-conquer approach to extract comprehensive protocol formats from RFCs, covering message types, field names, field types, independent field constraints, and dependent constraints, enabling more accurate protocol modeling and testing.
LLMIF~\cite{LLMIF} also leverages LLMs to extract formats from documents for test case generation. 
But unlike \sysname, which queries LLMs only when an inconsistency is detected between the protocol parser’s output and the expected format, LLMIF queries LLMs for every test case.
Additionally, LLMIF is tailored to Zigbee protocols and relies heavily on Zigbee-specific structures like clusters, commands, and ZCL frames.
This makes it challenging to adapt to general protocols (lack of these structures) without significant modification. 
In contrast, \sysname is designed for general protocol testing.
\vspace{-2mm}
\section{Conclusion}\label{sec:conclusion}
This work proposes \sysname to automatically validate network protocol implementations (in various programming languages) using protocol format specifications extracted from RFC documents. 
\sysname accurately extracts formats through a divide-and-conquer approach and thoroughly tests protocol parsers with fine-grained testing. 
\sysname also supports traceable inconsistency identification, allowing each inconsistency to be traced back to the original document section for accurate diagnosis.
Our experiments show that \sysname extracts protocol formats precisely, outperforming ChatAFL. \sysname detects 69 bugs in total, demonstrating the potential for automated software validation from natural language specifications.

\section*{Acknowledgment}
We thank all the anonymous reviewers for the insightful feedback.
We are grateful to the Center for AI Safety for providing computational resources. This work was funded in part by the National Science Foundation (NSF) Awards SHF-1901242, SHF-1910300, Proto-OKN 2333736, IIS-2416835, DARPA VSPELLS - HR001120S0058, and ONR N00014-23-1-2081. Any opinions, findings and conclusions or recommendations expressed in this material are those of the authors and do not necessarily reflect the views of the sponsors.

\bibliographystyle{ACM-Reference-Format}
\bibliography{sample-base}
\end{document}